\documentclass[10pt,twocolumn,letterpaper]{article}

\usepackage{iccv}
\usepackage{times}
\usepackage{epsfig}
\usepackage{graphicx}
\usepackage{amsmath}
\usepackage{amssymb}

\usepackage{wrapfig}

\usepackage[pagebackref=true,breaklinks=true,letterpaper=true,colorlinks,bookmarks=false]{hyperref}

\iccvfinalcopy %

\def\iccvPaperID{7968} %

\ificcvfinal\fi

\begin{document}

\title{Differentiable Surface Rendering via Non-Differentiable Sampling}

\author{Forrester Cole$^{1}$
\qquad
Kyle Genova$^{1}$
\qquad
Avneesh Sud$^{1}$
\qquad
Daniel Vlasic$^{1}$
\qquad
Zhoutong Zhang$^{1,2}$\vspace{0.1cm}\\
$^1$Google Research \qquad \qquad $^2$MIT\\
}

\maketitle

\ificcvfinal\thispagestyle{empty}\fi

\begin{abstract}
We present a method for differentiable rendering of 3D surfaces that supports both explicit and implicit representations, provides derivatives at occlusion boundaries, and is fast and simple to implement. The method first samples the surface using non-differentiable rasterization, then applies differentiable, depth-aware point splatting to produce the final image. Our approach requires no differentiable meshing or rasterization steps, making it efficient for large 3D models and applicable to isosurfaces extracted from implicit surface definitions. We demonstrate the effectiveness of our method for implicit-, mesh-, and parametric-surface-based inverse rendering and neural-network training applications. 
In particular, we show for the first time efficient, differentiable rendering of an isosurface extracted from a neural radiance field (NeRF), and demonstrate surface-based, rather than volume-based, rendering of a NeRF.
\end{abstract}
\vspace{-1em}

\section{Introduction}

Computing the derivatives of rendered surfaces with respect to the underlying scene parameters is of increasing interest in graphics, vision, and machine learning. Triangle meshes are the predominant shape representation in many industries, but mesh-based derivatives are undefined at occlusions or when changing topology. As a result, volumetric representations have risen in prominence for computer vision applications, notably Neural Radiance Fields or NeRF~\cite{nerf}. So far, these volumetric shape representations have been rendered using volume rendering. Volume rendering is naturally differentiable, but is expensive and unnecessary if the underlying shape can be represented well by a surface.

This paper proposes a method to render both explicit (e.g., mesh) and implicit (e.g., isosurface) representations and produce accurate, smooth derivatives, including at occlusion boundaries. Our method uses a non-differentiable rasterization step to sample the surface and resolve occlusions, then splats the samples using a depth-aware, differentiable splatting operation. Because the sampling operation need not be differentiable, any conventional surface extraction and rasterization method (e.g., Marching Cubes~\cite{marchingcubes}) may be used. The splats provide smooth derivatives of the image w.r.t. the surface at occlusion boundaries. Splatting is performed on a fixed-size pixel grid and is easily expressed using automatic-differentiation, avoiding the need for custom gradients. Since no custom gradients are needed, both forward- and reverse-mode differentiation are immediately supported. We term this method \emph{rasterize-then-splat} (RtS).

\begin{figure}
    \centering
    \includegraphics[width=\columnwidth]{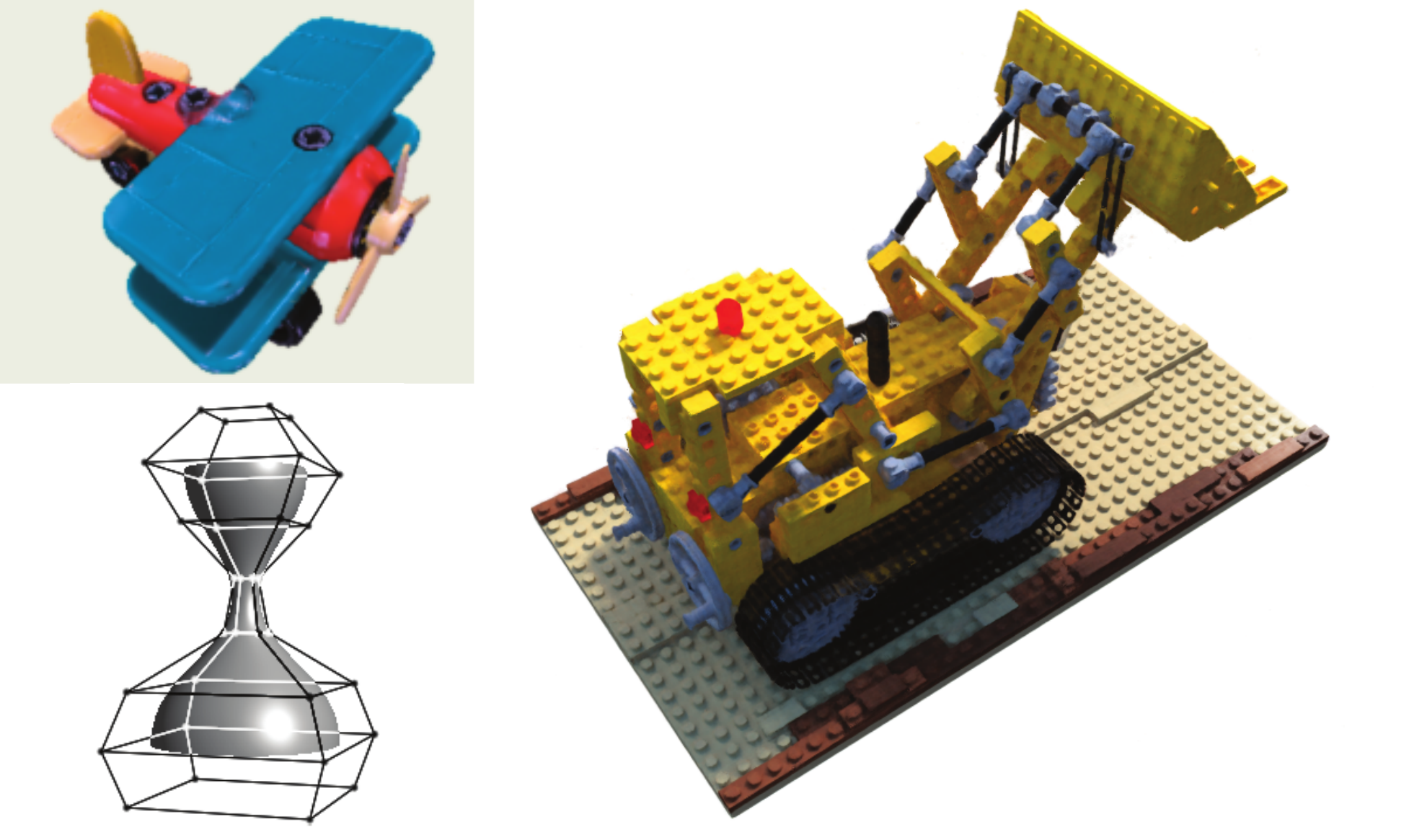}
    \caption{Our method provides efficient, differentiable rendering for explicit and implicit surface representations. Examples include a textured triangle mesh (YCB toy airplane~\cite{YCB}), a cubic B-spline surface, and an isosurface of a density volume (Lego from NeRF~\cite{nerf}). The Lego is rendered by turning a pretrained NeRF into a \emph{surface light field}. Since surface light fields only require one evaluation per pixel, we achieve a $128\times$ speed up for rendering compared with the original NeRF. 
    }
    \label{fig:teaser}
    \vspace{-1em}
\end{figure}

In between the rasterization and splatting steps, the surface samples may be shaded by any differentiable function evaluated on a rasterized image buffer -- not the original surface -- using deferred shading~\cite{deferredshading}. Since the complexity of the shading and splatting computation is bounded by the number of pixels, not the complexity of the surface, RtS is able to scale to highly detailed scenes.

One example of a differentiable shading function is a NeRF network: given a position in space and a viewing direction, it outputs the corresponding radiance. While NeRF is trained using volume rendering, our method can convert a pretrained NeRF into a \emph{surface light field}~\cite{miller98,slf}, removing the need for expensive raymarching. We represent the surface as an isosurface of the density field extracted from a pretrained NeRF, shade it with the NeRF color prediction branch, and jointly finetune the NeRF network and the density field. 
The resulting optimized surface and surface light field matches the original NeRF network in rendering quality (within $0.3$ PSNR) but requires only a single network evaluation per pixel, producing a 128$\times$ speedup (Fig.~\ref{fig:teaser}). 

We further demonstrate that RtS provides high-quality derivatives for inverse rendering of meshes and parametric surfaces, while remaining simple to implement. An implementation of RtS for mesh-based rendering is provided as part of TensorFlow Graphics\footnote{https://www.tensorflow.org/graphics}.

\begin{figure*}
    \centering
    \includegraphics[width=2\columnwidth]{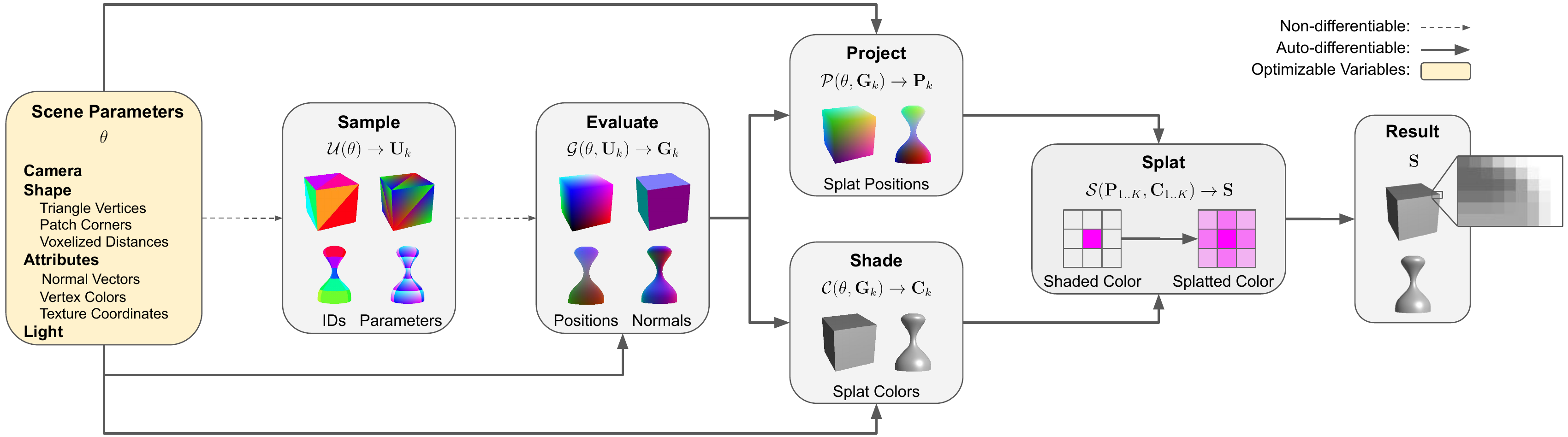}
    \caption{\textbf{Rasterize-then-splat system.} Scene parameters $\mathbf{\theta}$ are first passed through a sampling function $\mathcal{U}$, yielding per-layer screen-space parameter buffers $\mathbf{U}_k$. These buffers are non-differentiable and computed only in the forward pass (dotted arrows). Evaluation function $\mathcal{G}$ uses them to interpolate shape attributes, such as positions and normals, into G-buffers $\mathbf{G}_k$. Automatic differentiation may be applied to $\mathcal{G}$ and downstream functions (solid arrows).  The evaluated attributes are combined to compute splat colors $\mathbf{C}_k$ via deferred shading function $\mathcal{C}$ and the corresponding screen-space positions $\mathbf{P}_k$ via projection function $\mathcal{P}$. Function $\mathcal{S}$ then splats the shaded colors the corresponding pixel locations using a 3x3 kernel to produce the final result $\mathbf{S}$.}
    \label{fig:system_diagram}
    \vspace{-0.5em}
\end{figure*}

\section{Related Work}
\label{sec:related}

Early differentiable rendering explored the derivatives of specialized shape parameterizations (e.g., a morphable model~\cite{blanz1999morphable} or a heightfield~\cite{smelyansky2002dramatic,jalobeanu2004modeling,BarronTPAMI2015}). Recent work has focused on general 3D triangle meshes and implicit representations such as signed-distance fields and volumes.

\subsection{Rendering Triangle Meshes}

When rendering triangle meshes, topology is assumed to be constant. The remaining major challenge for computing derivatives is handling of occlusion boundaries. Previous work falls into four main categories: 

\noindent\textbf{Gradient Replacement.} Methods such as \cite{nmr,opendr,dirt} use heuristics to define smooth derivatives for the mesh rendering while leaving the forward rendering unchanged. OpenDR~\cite{opendr} and DiRT~\cite{dirt} use image-space filtering operations to find the derivatives, while Neural Mesh Renderer (NMR)~\cite{nmr} defines a special, per-triangle derivative function. A differentiable version of surface splatting~\cite{zwicker2001surface} is proposed in~\cite{dss} with a modified gradient function. These approaches do not easily support textures or complex shading models, and in some cases produce convergence problems due to the mismatch between the rendering and its gradient.
Custom gradient functions are implemented only for the Jacobian-vector product necessary for gradient backpropagation, and new, additional functions are necessary to support forwards-mode or higher-order derivatives.

\noindent\textbf{Edge Sampling.} Redner~\cite{redner}, nvdiffrast~\cite{nvdiffrast}, DEODR~\cite{deodr}, and others \cite{gargallo2007minimizing,delaunoy2011gradient} explicitly sample the occluding edges of the shape to compute derivatives. They require shape processing to find and sample the edges, so the cost of computing derivatives grows with the number of edges in the mesh. Nvdiffrast mitigates the cost using tightly optimized CUDA kernels, however their code requires specific GPU hardware and is not easy to alter for new systems.
RtS can be implemented without any shape processing or custom derivative code, and the cost of the differentiable sections is independent of the size of the mesh. 

\noindent\textbf{Reparameterizing the Rendering Integral.} When performing Monte-Carlo path tracing, occlusion discontinuities may be handled using reparameterizations of the rendering equations \cite{mitsuba2, bangaru2020unbiased}. These methods are related to ours in that they choose surface samples without explicit sampling of occlusion boundaries. However, these methods apply only in the context of path tracing, while RtS supports simple shading and rasterization. 

\noindent\textbf{Forward Rendering with Smooth Gradients.} Similar to our approach, Soft Rasterizer and related methods~\cite{softras,pytorch3d,dibr}, as well as the differentiable visibility method of \cite{gaussianballs} change the forward rendering process such that its gradient is smooth by construction. Unlike Soft Rasterizer, RtS does not require costly closest-point queries or mesh processing.

\subsection{Surface Splatting}

Surface splatting~\cite{zwicker2001surface} treats the surface as a point cloud and renders disk primitives at each point that overlap to create a continuous surface. Splatting has been adapted for differentiable rendering of 3D shapes~\cite{dss} and forward-warping of images~\cite{niklaus2020softmax}. Similar to splats, differentiable visibility using 3D volumetric primitives has also been explored~\cite{gaussianballs}. Compared to these approaches, our method uses a true surface representation as the underlying geometry, rather than a point set, and resamples the splats at each frame, avoiding issues with under- or over-sampling of splats as optimization proceeds.

\subsection{Rendering Implicit Surfaces}

Implicit surface representations such as signed-distance fields naturally handle topological changes, but rendering still requires explicit handling of occlusion boundaries. In recent work, an occlusion mask is sometimes assumed to be provided by the user~\cite{yariv2020multiview,niemeyer2020dvr}, or computed by finding the nearest grazing point on a ray that hits the background~\cite{liu2020dist}. Neither method handles self-occlusion, which is the only type of occlusion in walkthrough-style scenes (Fig.~\ref{fig:nerf_results}).

Volume rendering~\cite{elvins1992survey,nerf} provides smooth derivatives at occlusions, including self-occlusions, but requires expensive ray marching to find the surface. Marching Cubes (MC) isosurface extraction~\cite{marchingcubes} may be used to convert the volume into a surface for optimization, but this process is not naturally differentiable~\cite{liao2018deep}. Our method extracts and rasterizes the isosurface in a single non-differentiable step, then computes derivatives in image-space, avoiding the singularity in the MC derivative.

Most related to RtS is MeshSDF~\cite{meshsdf}, which also uses non-differentiable sampling of the implicit surface, followed by differentiable occlusion testing using NMR~\cite{nmr}. However, MeshSDF defines a custom derivative using the normal of the SDF, a technique that holds for true SDFs but not for general isosurfaces. Further, MeshSDF demonstrates only a neural representation of the surface, whereas our method allows isosurfaces parameterized by a grid or any other differentiable function.

\section{Method}

The rasterize-then-splat method consists of three steps: rasterization of the surface (Sec.~\ref{sec:rasterization}), shading the surface samples (Sec.~\ref{sec:shading}), and multi-layer splatting (Sec.~\ref{sec:splatting}). All derivatives are produced using automatic differentiation, so the implementer only needs to write the forward rendering computation (Fig.~\ref{fig:system_diagram}).

\subsection{Rasterization via Non-Differentiable Sampling}
\label{sec:rasterization}

Rasterization can be expressed as a function that takes scene parameters $\theta$ containing geometry attributes such as position, normal, or texture coordinates, as well as camera parameters, and produces screen-space geometry buffers (or \emph{G-buffers}~\cite{saito:1990:CRDSy}) $\mathbf{G}_{k \in 1 .. K}$ containing interpolated attributes at the $K$ closest ray intersections to the camera. To make this process both differentiable and efficient, we divide rasterization into two stages: a \emph{sampling} function $\mathcal{U}(\theta) \rightarrow \mathbf{U}_k$ that produces non-differentiable surface parameters $\mathbf{U}_k$, and an \emph{evaluation} function $\mathcal{G}(\theta, \mathbf{U}_k) \rightarrow \mathbf{G}_k$ that produces the G-buffers. The necessary parameters vary with the surface type (see below). 

Evaluation of surface attributes given surface parameters is typically a straightforward interpolation operation, so $\mathcal{G}$ can be easily expressed in an automatic-differentiation framework. The difficult and computationally-intensive operation is the sampling function $\mathcal{U}$ that finds the intersections of the surface with the camera rays. However, since we are not interested in derivatives w.r.t. the sampling pattern itself, $\mathcal{U}$ may act as a non-differentiable ``oracle'' that finds the intersections but produces no derivatives for them. 

Below we give concrete examples of $\mathcal{U}$ and $\mathcal{G}$ for triangle meshes, parametric surfaces, and implicit surfaces.

\subsubsection{Triangle Meshes}

For triangle meshes, the parameters $\mathbf{U}_k$ consist of per-pixel triangle indices $\mathbf{T}_k$ and the (perspective-correct) barycentric coordinates $\mathbf{B}_k$ of the pixel inside the corresponding triangle. The sampling function $\mathcal{U}$ can compute these values extremely efficiently with conventional Z-buffer graphics processing, using depth peeling~\cite{everitt2001interactive} to retrieve multiple intersections per pixel. The evaluation function $\mathcal{G}$ simply looks up the three vertex attributes for each pixel using $\mathbf{T}$, then interpolates them using $\mathbf{B}$.

\subsubsection{Parametric Surfaces}

Bicubic regular B-spline surfaces~\cite{Catmull1978RecursivelyGB} are a type of smooth parametric surface, a representation that so far has not supported differentiable rendering. Efficient rasterization of these surfaces is achieved by subdividing rectangular patches until the resulting facets are smaller than a pixel, complicating the propagation of derivatives. We avoid this difficulty with the non-differentiable sampling function $\mathcal{U}$ that returns per-pixel patch indices and patch parameters. The evaluation function $\mathcal{G}$ then interpolates the patch vertex attributes using the parameters and the B-spline basis matrix (Equation 1 in \cite{Catmull1978RecursivelyGB}). This approach can be extended to all parametric surfaces with a closed form evaluation expression, such as Catmull-Clark subdivision surfaces~\cite{stam1998} and B\'{e}zier surfaces.

\subsubsection{Implicit Surfaces}
\label{implicit_method}

We treat implicit surfaces as isocontours of a function $f_{\theta}: \mathbb{R}^3 \rightarrow \mathbb{R}$ over 3D space. Unlike meshes or spline patches, implicit surfaces do not have a natural parameterization. We choose a parameterization based on the triangulation of the isosurface provided by the Marching Cubes~\cite{marchingcubes} algorithm. The parameters $\mathbf{U}_k$ are 9-D vectors, consisting of 6 lattice indices $v_{1..6}$ defining the 3 edges that cross the isosurface, and 3 triangle barycentric coordinates $\beta_{1..3}$. 

\begin{wrapfigure}{r}{0.35\columnwidth}
    \centering
    \hspace{-0.1\columnwidth}\includegraphics[width=0.45\columnwidth]{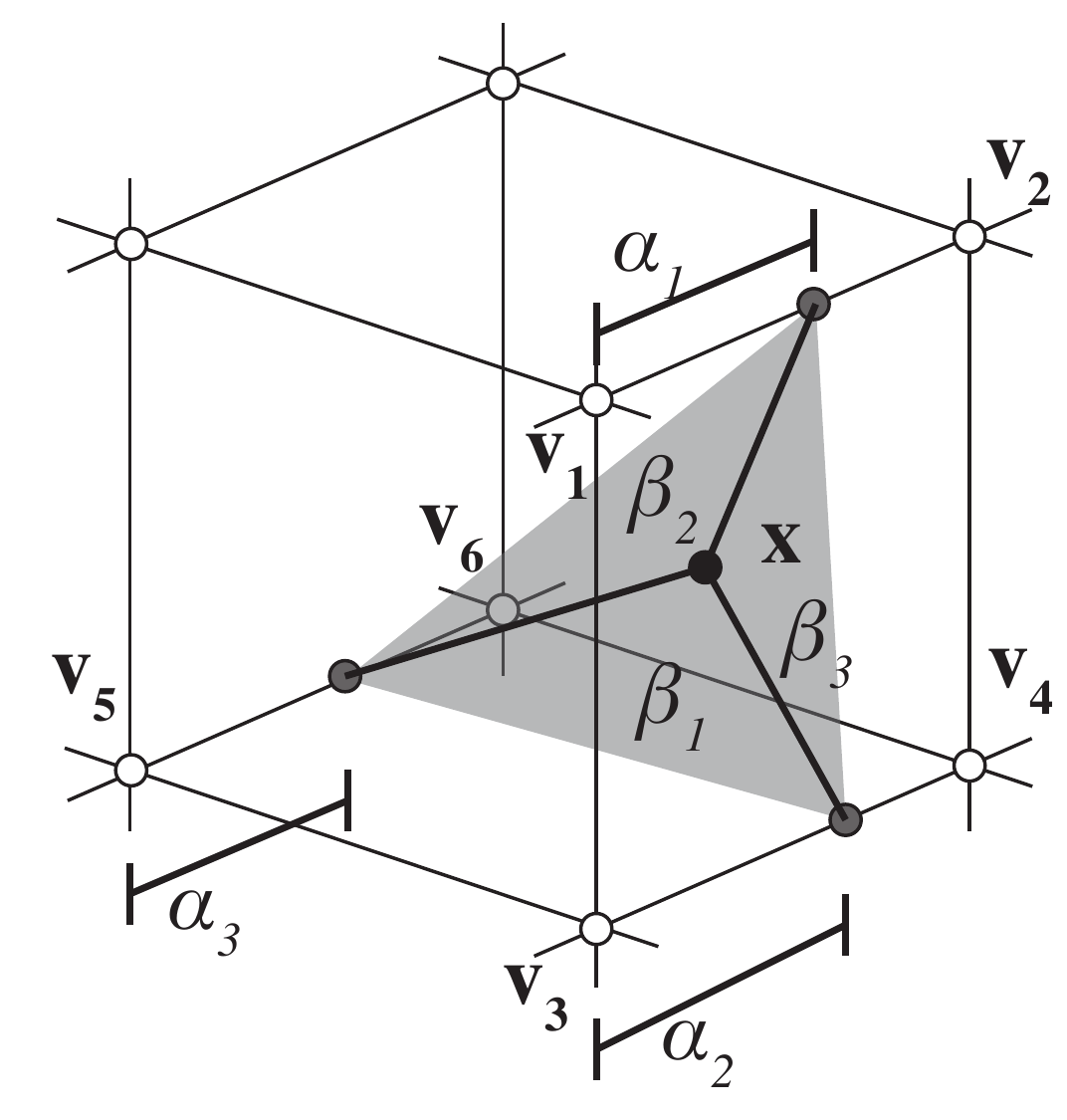}
    \label{fig:voxel_parameterization}
    \vspace{-2em}
\end{wrapfigure}

The evaluation function $\mathcal{G}$ evaluates $f$ at the $v_{1..6}$ (simply looking up their values if $f$ is already defined on a grid), interpolates along the edges to find the coefficients $\alpha_{1..3}$ that define the triangle vertices, then interpolates the vertices using $\beta_{1..3}$ to produce the surface point $\mathbf{x}$. Critically, this scheme hides the complex topological rules of Marching Cubes in the non-differentiable function $\mathcal{U}$, removing the need to store or differentiate through the full mesh topology ~\cite{liao2018deep}. 

Note that while the Marching Cubes algorithm is not itself differentiable due to the singularity when neighboring grid values are nearly identical~\cite{liao2018deep, meshsdf}, our procedure side-steps this issue by evaluating the surface only where the derivative is well-defined. We remove samples from nearly identical grid cells in the sampling function $\mathcal{U}$.

\subsection{Shading}
\label{sec:shading}

The G-buffers $\mathbf{G}_k$ contain various surface attributes depending on the shading required. Any shading function $\mathcal{C}$ that can be expressed as a deferred shading operation~\cite{deferredshading} can be applied. For a texture-mapped mesh (Fig.~\ref{fig:airplane_optimization}), each pixel in $\mathbf{G}_k$ contains a 3D position, a 3D surface normal and 2D texture coordinates. For parametric surface rendering (Fig.~\ref{fig:spline_results}) and implicit surface rendering using a NeRF shader (Fig.~\ref{fig:nerf_results}), $\mathbf{G}_k$ contains only 3D world-space positions. The output of the shading step is a set of RGBA buffers $\mathbf{C}_k$.

\subsection{Depth-Aware Splatting}
\label{sec:splatting}

The shaded colors $\mathbf{C}_k$ have derivatives w.r.t. the surface attributes, but because they were produced using point sampling, they do \emph{not} have derivatives w.r.t. occlusion boundaries. To produce smooth derivatives at occlusions, the splatting function $\mathcal{S}$ converts each rasterized surface point into a splat, centered at the corresponding pixel in $\mathbf{P}_k$ and colored by the corresponding shaded color in $\mathbf{C}_k$. In order to handle splat overlap at occlusion boundaries, we introduce a multi-layer accumulation strategy for the splats based on depth (Sec.~\ref{sec:multilayersplat}) that provides superior accuracy for occlusions and disocclusions  (Sec.~\ref{sec:analysisofderivatives}).

Though a splat is always centered on a pixel, the position of the splat must be computed using the surface definition in order for derivatives to flow from the image back to the surface. The splat positions are defined by an additional G-buffer $\mathbf{P}_k$, which contains the screen-space $xyz$ positions of each surface sample.
$\mathbf{P}_k$ may be computed by rendering a G-buffer of object-space $xyz$ positions (Sec.~\ref{sec:rasterization}), then applying the camera view and projection transformation at each pixel.

\subsubsection{Single-Layer Splatting}

The splat kernel is defined by a Gaussian with narrow variance. If $\mathbf{p}$ is the center position of a single splat, the weight of the splat at a nearby pixel $\mathbf{q}$ is:
\begin{equation}
    w_{\mathbf{p}}(\mathbf{q}) = \frac{1+\epsilon}{\mathbf{W}_\mathbf{p}} \; \mathtt{exp}\left(\frac{-\left\lVert\mathbf{q} - \mathbf{p}\right\rVert^2_2}{2\sigma^2}\right)
\label{eq:splatdef}
\end{equation}
where $\epsilon$ is an small adjustment factor, and $\mathbf{W}_\mathbf{p}$ is a normalization factor computed from the sum of all weights in a 3x3 neighborhood. By setting $\sigma=0.5$, we have:
\begin{equation}
    \mathbf{W}_\mathbf{p} = \sum_{i=-1}^{1}\sum_{j=-1}^{1}\mathtt{exp}\left(-2\left\lVert\left\lfloor{\mathbf{p}}\right\rfloor+(i,j) - \mathbf{p}\right\rVert^2_2\right).
\end{equation}

The final color $\mathbf{s}_\mathbf{q}$ at pixel $\mathbf{q}$ is then the weighted sum of the shaded colors $\mathbf{c}_{\mathbf{r}}$ of the neighboring pixels $\mathbf{r} \in N_\mathbf{q}$ divided by the accumulated weights:
\begin{equation}
    \mathbf{s}_\mathbf{q} = \left( \sum_{\mathbf{r} \in N_\mathbf{q}} w_\mathbf{r}(\mathbf{q}) \; \mathbf{c}_{\mathbf{r}} \right) \bigg/ \mathtt{max} \left( 1, \sum_{\mathbf{r} \in N_\mathbf{q}} w_\mathbf{r}(\mathbf{q}) \right)
\label{eq:splatnorm}
\end{equation}
where the normalization factor has a floor of $1$ to handle boundaries where the weights in $N_\mathbf{q}$ may sum to $<1.0$. Due to the adjustment factor $\epsilon=0.05$, a full 3x3 neighborhood of weights always sum to $>1.0$ (see below).

\paragraph{Importance of Normalization.} 

The need for the adjustment factor $\epsilon$ in Eq.~\ref{eq:splatdef} and the additional normalization in Eq.~\ref{eq:splatnorm} is subtle; 
the splats are resampled at exactly pixel rate every frame, so normalization by the accumulated weights of neighboring splats as in \cite{zwicker2001surface} is not necessary for forward rendering. The derivatives of $\mathbf{s_q}$, however, do not account for resampling, and do need to be normalized by the accumulated weights in order to match the forward pass. Since we want to allow the accumulated weights to sum $<1$ at boundaries with the background, we add $\epsilon$ to ensure the normalization always occurs for interior splats.

\begin{figure*}[h]
    \centering
    \includegraphics[width=\linewidth]{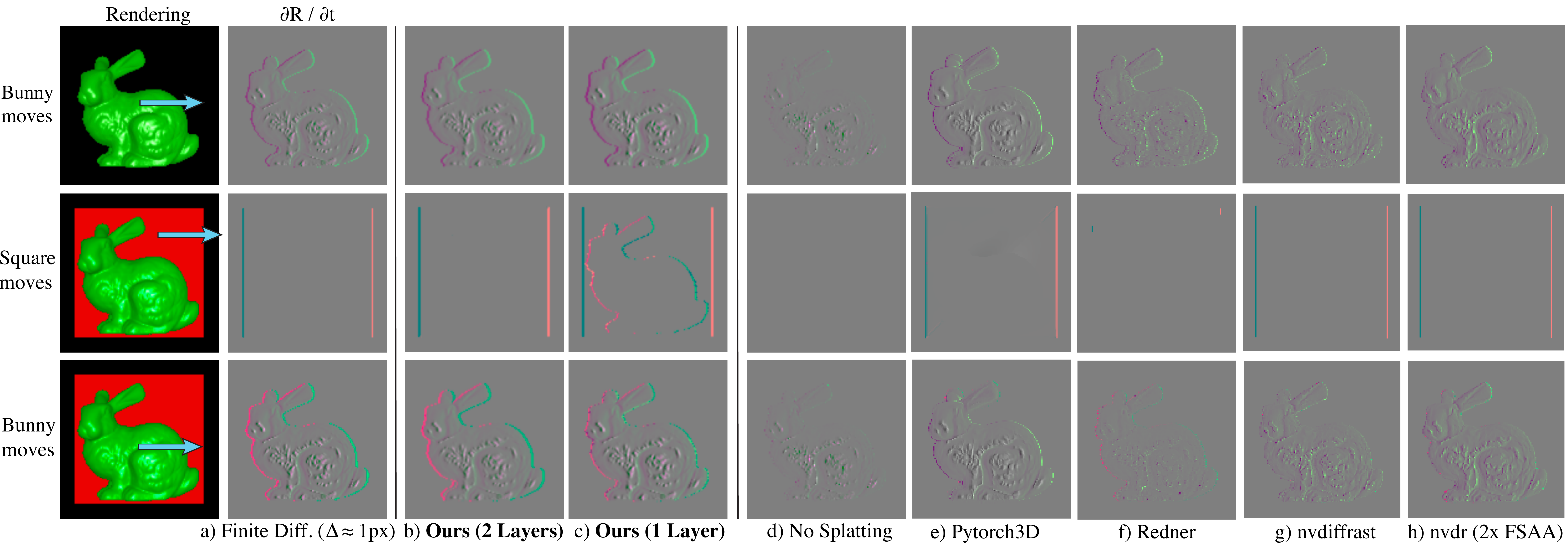}
    \caption{\textbf{Visualization of rendering derivatives for our and baseline methods.} From top to bottom: the green bunny moves to the right on black background, the red cube moves behind the bunny, and the bunny moves in front of the cube. Finite differences (a) can be treated as ground-truth in this case. Our method with 2 layers (b) closely matches the finite difference result, while ours with 1 layer (c) produces spurious derivatives when the cube moves behind the bunny. 
    Baseline methods (d-h) shown for comparison. See Sec~\ref{sec:analysisofderivatives} for analysis.
    }
    \label{fig:bunny_cube_derivatives}
\end{figure*}

\subsubsection{Multi-Layer Splatting}
\label{sec:multilayersplat}

\begin{figure}[b]
    \centering
    \includegraphics[width=\columnwidth]{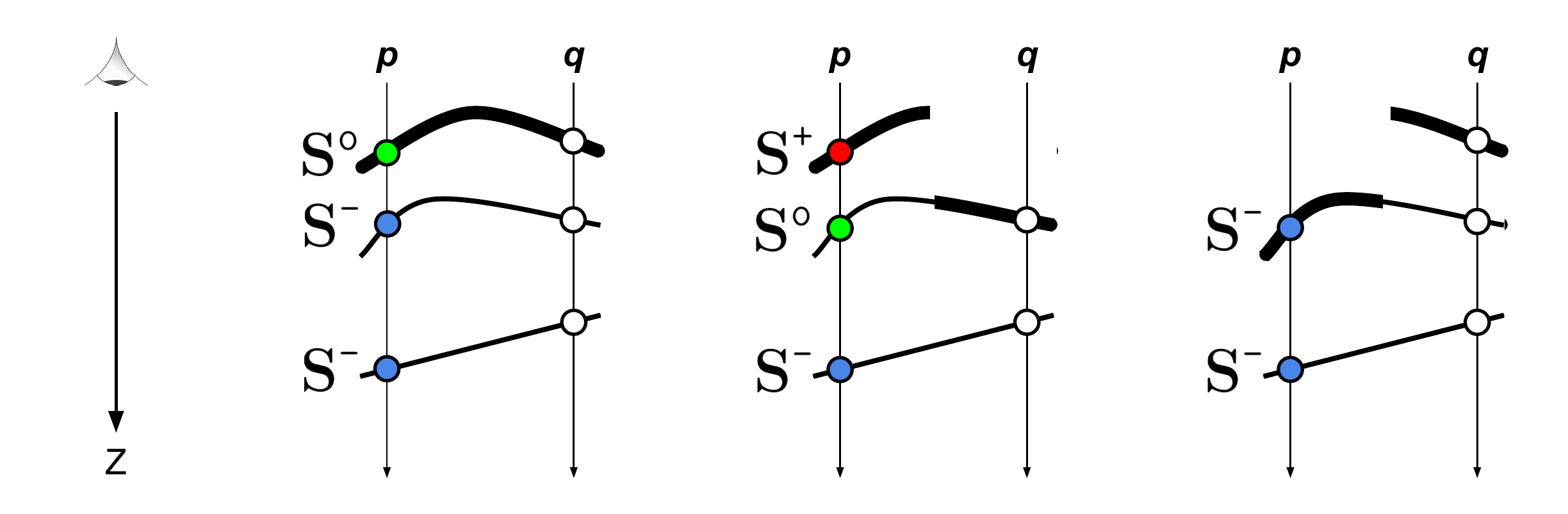}
    \caption{\textbf{Multi-layer splat accumulation.} When the splat at $\mathbf{p}$ is being applied to the pixel $\mathbf{q}$, it contributes to one of the three accumulation buffers: $\mathbf{S}^\mathtt{o}$ (green) when $\mathbf{p}$ is coincident with the visible surface at $\mathbf{q}$ (thick curve), $\mathbf{S}^\texttt{+}$ (red) when $\mathbf{p}$ is in front, or $\mathbf{S}^\texttt{-}$ (blue) when $\mathbf{p}$ is behind.  Different cases arise depending on whether there is an occlusion boundary between $\mathbf{p}$ and $\mathbf{q}$.  We ignore rare cases of multiple coincident occlusion boundaries.
    }
    \label{fig:multilayer_splat}
    \vspace{-0.5em}
\end{figure}

Single-layer splatting treats all splats as existing at the same depth and ignores occlusions, producing spurious derivatives for occluded objects (Fig.~\ref{fig:bunny_cube_derivatives}c). Instead, depending on a splat's relation to the visible surface at a target pixel, it should either occlude the pixel, be occluded itself, or be accumulated as in Eq.~\ref{eq:splatnorm}.

Our solution is to render multiple layers of G-buffers, and maintain three accumulation buffers during the splatting process: $\mathbf{S}^\texttt{+}$ for splats occluding the target pixel, $\mathbf{S}^\texttt{-}$ for occluded splats, and $\mathbf{S}^\texttt{o}$ for splats at the same depth as the target pixel.  When applying a splat centered at $\mathbf{p}$ to a pixel $\mathbf{q}$, weighted colors and weights are accumulated into exactly one of the three buffers (Fig.~\ref{fig:multilayer_splat}).

To determine whether the splat lies in front on, behind, or coincident with the pixel, we propose a simple heuristic that is more robust than picking a fixed depth threshold.  We pair up the multi-layer surface intersections at $\mathbf{p}$ with the closest intersections at $\mathbf{q}$ in depth. The $\mathbf{p}$ layer paired with the front-most $\mathbf{q}$ layer is assigned to $\mathbf{S}^\texttt{o}$, layers in front of it (if any) to $\mathbf{S}^\texttt{+}$, and the rest to $\mathbf{S}^\texttt{-}$.

Once all splats are rendered, buffers are separately normalized following Eq.~\ref{eq:splatnorm} and composited in $\mathbf{S}^\texttt{-}, \mathbf{S}^\texttt{o}, \mathbf{S}^\texttt{+}$ order using over-compositing~\cite{porterandduff} to produce the final result $\mathbf{S}$. This scheme correctly handles occlusions between the first and second layers of the surface (Fig.~\ref{fig:bunny_cube_derivatives}d). 

\section{Results and Evaluation}

\subsection{Analysis of Derivatives}
\label{sec:analysisofderivatives}

Fig.~\ref{fig:bunny_cube_derivatives} visualizes the image derivatives for a green bunny superimposed on a black background and on a diffuse red square. The derivative shown is $\partial \mathbf{S}/ \partial t$, where $t$ is the translation of either the bunny or the square. Finite differences (Fig.~\ref{fig:bunny_cube_derivatives}a) provide a ``ground-truth,'' since $\Delta t$ can be chosen to produce $\approx 1$ pixel motion in this case.
Multi-layer splatting (Fig.~\ref{fig:bunny_cube_derivatives}b) produces derivatives that closely resemble the finite difference result. Single-layer splatting (Fig.~\ref{fig:bunny_cube_derivatives}c) provides derivatives at occlusion boundaries, but confuses self-occlusions: when the red square moves behind the bunny (middle row), single-layer splatting produces a spurious derivative around the bunny's outline. 

Fig.~\ref{fig:bunny_cube_derivatives}(d-h) show baseline methods for comparison.
Differentiable rasterization without splatting (d) provides derivatives in the interior of the shape, but not at the occlusion boundaries. PyTorch3D~\cite{pytorch3d} (e) produces inaccurate derivatives for self-occlusions (bottom row). Redner~\cite{redner} (f) better handles self-occlusions, but may miss edges due to sampling (middle row). nvdiffrast~\cite{nvdiffrast} (g,h) relies on pixel sampling to find edges and so misses sub-pixel edges as exist along the bunny outline. FSAA (h) improves the result but does not solve the problem completely. See supplemental material for the parameters used for these comparisons.

While the result $\mathbf{S}$ looks as if $\mathbf{C}_k$ were simply blurred slightly, blurring $\mathbf{S}$ is not sufficient to produce non-zero derivatives w.r.t. the surface at occlusions. As shown in Fig.~\ref{fig:bunny_cube_derivatives}b, rasterization without splatting produces zero derivative at occlusion boundaries, so any blur following rasterization will also produce zero derivative. 

\textbf{Effect of blur.} The blur applied by our method is slight ($\sigma$ = 0.5px), though not invisible. To analyze whether this blur affects optimization when image sharpness is important, we repeat the texture optimization experiment from Fig. 6 of \cite{nvdiffrast}, which optimizes a textured sphere to match target images. With mipmapping on, a blur of $\sigma=0.5$ reduces PSNR by 1.5\% from 33.7 to 33.2. With mipmapping off, the blur \emph{increases} PSNR from 25.4 to 28.3, likely due to the slight anti-aliasing effect of the blur.

\subsection{Pose Estimation}

A common use-case for differentiable rendering is to estimate the pose of a known object given one or more images of that object. Compared with previous work, RtS is particularly suitable for this task because its runtime increases slowly with mesh complexity, and it supports more sophisticated optimizers than gradient descent.

\begin{figure}[t]
    \centering
    \begin{tabular}{cccccc}
         \includegraphics[height=0.16\columnwidth]{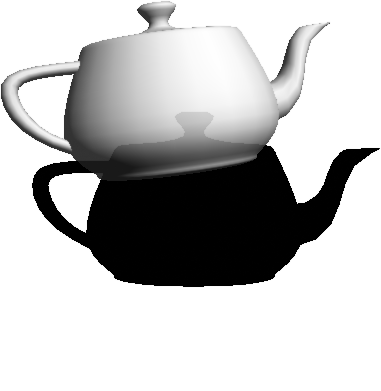} & 
         \includegraphics[height=0.16\columnwidth]{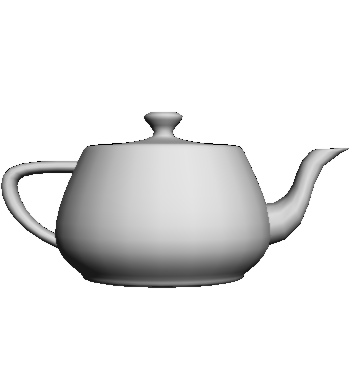} & 
         \includegraphics[height=0.19\columnwidth]{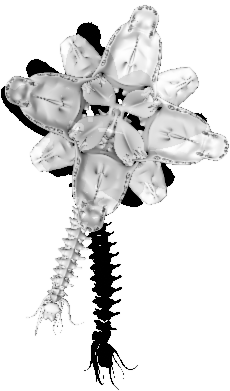} & 
         \includegraphics[height=0.19\columnwidth]{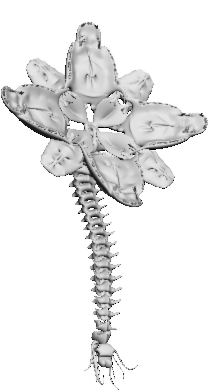} &
         \includegraphics[height=0.19\columnwidth]{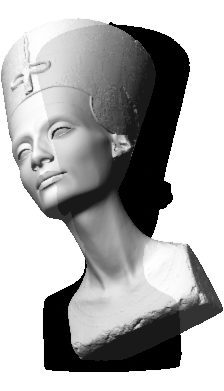} &
         \includegraphics[height=0.19\columnwidth]{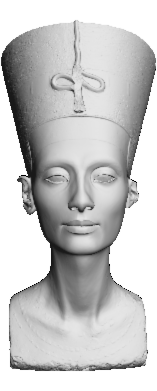} \\
    \end{tabular}
    \vspace{-1em}
    \caption{\textbf{Pose estimation from silhouettes.} Given a perturbed initial pose (left), three models are rotated and translated back to their original orientation (right) by following an $L_2$ pixel loss on the silhouette (black).}
    \label{fig:pose_estimation}
\end{figure}

\noindent\textbf{Performance Comparison.} RtS is fast for large meshes (Table~\ref{tab:pose_estimation}) as it uses a conventional forward rendering pass over the geometry, followed by image-space operations. PyTorch3D (based on Soft Rasterizer) requires spatial binning to achieve acceptable performance, and does not scale as well to large meshes. Redner~\cite{redner} similarly suffers due to the cost of sampling and processing the occluding edges. Nvdiffrast~\cite{nvdiffrast} achieves excellent performance at the cost of a complex, triangle-specific implementation. On a task of pose estimation from silhouettes (Fig.~\ref{fig:pose_estimation}), RtS achieves a speedup up to 20$\times$ over PyTorch3D and Redner for the Nefertiti ~\cite{keenan_meshes} mesh (2m faces), and smaller but significant speedups for the Teapot (2.5K faces). Our method performs within $2\times$ of Nvdiffrast without any custom CUDA kernels.

\begin{table}[b]
    \centering
    \begin{tabular}{l|ccccc}
         Triangles  & RtS & RtS-pose & P3D & Redner & Nvdr\\ \hline
         2.5K       & \underline{16} &\underline{16} & 21 & 240 & \bf{7}\\
         326K       & 18 & \underline{16} & 47 & 247 & \bf{9}\\
         2M         & 26 & \bf{17} & 588 & 306 & \underline{19}\\
    \end{tabular}
    \caption{\textbf{Pose estimation performance.} Milliseconds per iteration for alignment to silhouettes on V100. Methods compared are ``RtS'' (ours), ``RtS-pose`` (ours optimized for pose fitting), ``P3D'' (PyTorch3D ~\cite{pytorch3d}),  ``Redner'' ~\cite{redner}, and ``Nvdr'' (Nvdiffrast~\cite{nvdiffrast}).}
    \vspace{-1em}
    \label{tab:pose_estimation}
\end{table}

In the specific case of pose fitting to silhouettes, the sampling function $\mathcal{U}$ can return world-space positions directly, instead of triangle ids $\mathbf{T}$ and barycentric coordinates $\mathbf{B}$. Since the mesh itself is not changing, only the pose defined by the projection function $\mathcal{P}$, the world-space positions do not need to be differentiable and the evaluation function $\mathcal{G}$ can skip the potentially costly step of looking up vertex attributes given $\mathbf{T}$. This optimization (``RtS-pose'') removes the dependence on mesh complexity entirely from the differentiable components of the system, yielding performance largely independent of mesh complexity and faster than Nvdiffrast on the Nefertiti model.

\begin{figure}
    \centering
    \includegraphics[width=\columnwidth]{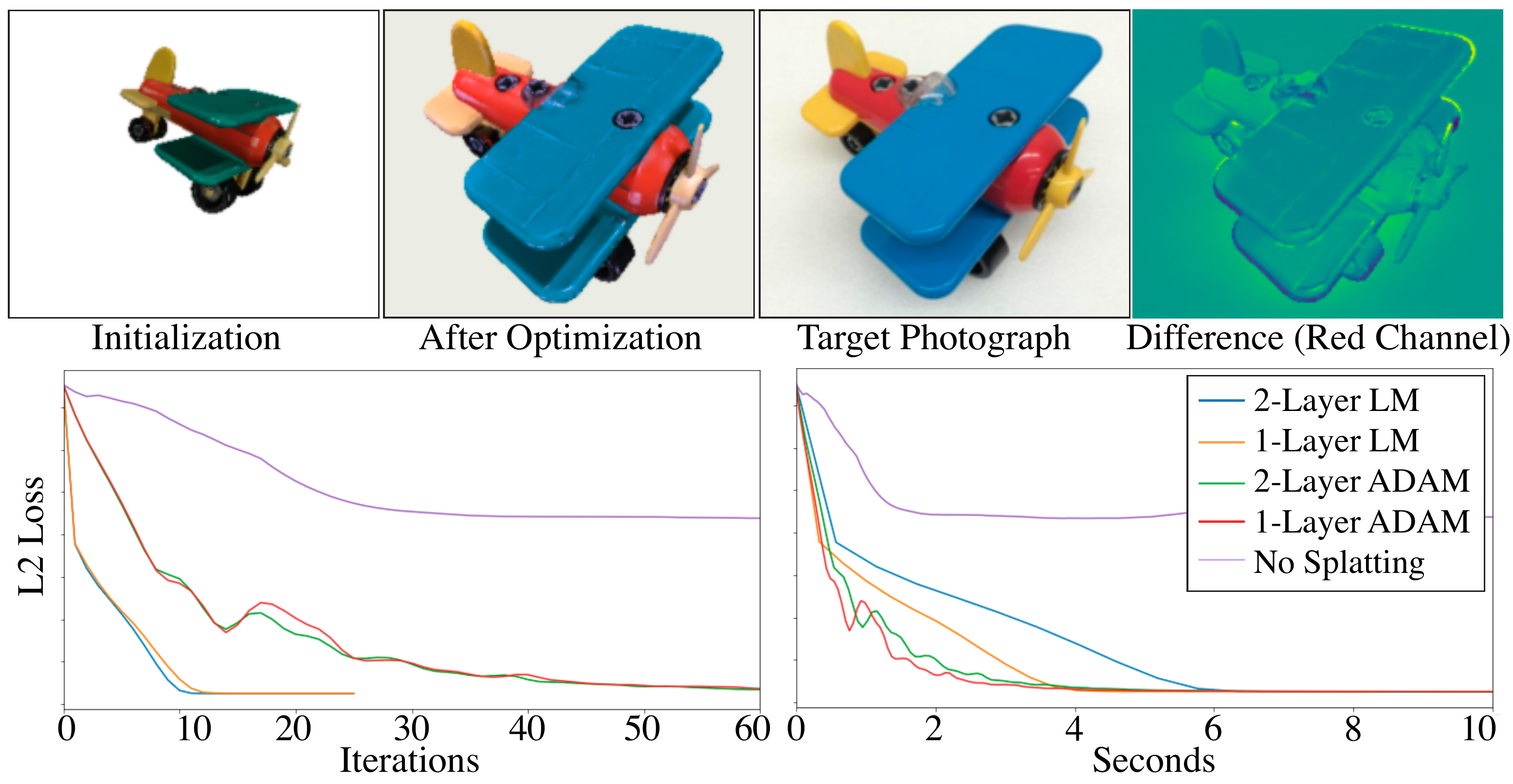}
    \caption{\textbf{LM vs. ADAM for pose estimation.} Starting from an initial camera pose, field of view, background color, and diffuse lighting (left), the rendering of the toy airplane~\cite{YCB} is optimized to match the photograph (middle right), resulting in a close pose match (right). Total iterations are plotted on left, wall time on right.  Levenberg-Marquardt~\cite{levenbergmarquardt} (LM) converges in fewer iterations and more smoothly than Adam~\cite{adam}, though total time to convergence is similar. Baseline rasterization (no splatting) does not converge to the correct solution. Multi-layer splatting has a limited effect for pose estimation.
    }
    \label{fig:airplane_optimization}
    \vspace{-1em}
\end{figure}

\noindent\textbf{Optimization with Levenberg-Marquardt.} Since RtS requires no custom gradient code, both forwards-mode and backwards-mode automatic differentiation can be applied. Pose estimation problems have fewer input variables (pose) than output variables (pixels), making forward-mode an efficient choice for computing the full Jacobian matrices required for optimization algorithms such as Levenberg-Marquadt~\cite{levenbergmarquardt}, which are prohibitively costly using backwards-mode differentiation. LM optimization provides robust convergence compared to Adam, though under our current implementation, the extra cost of LM means the two methods have similar total runtimes of $\approx4$ seconds to convergence (Fig.~\ref{fig:airplane_optimization}).

\subsection{Mesh Optimization}
\label{sec:mesh_optimization}

\begin{figure}[h]
    \centering
    \includegraphics[width=\columnwidth]{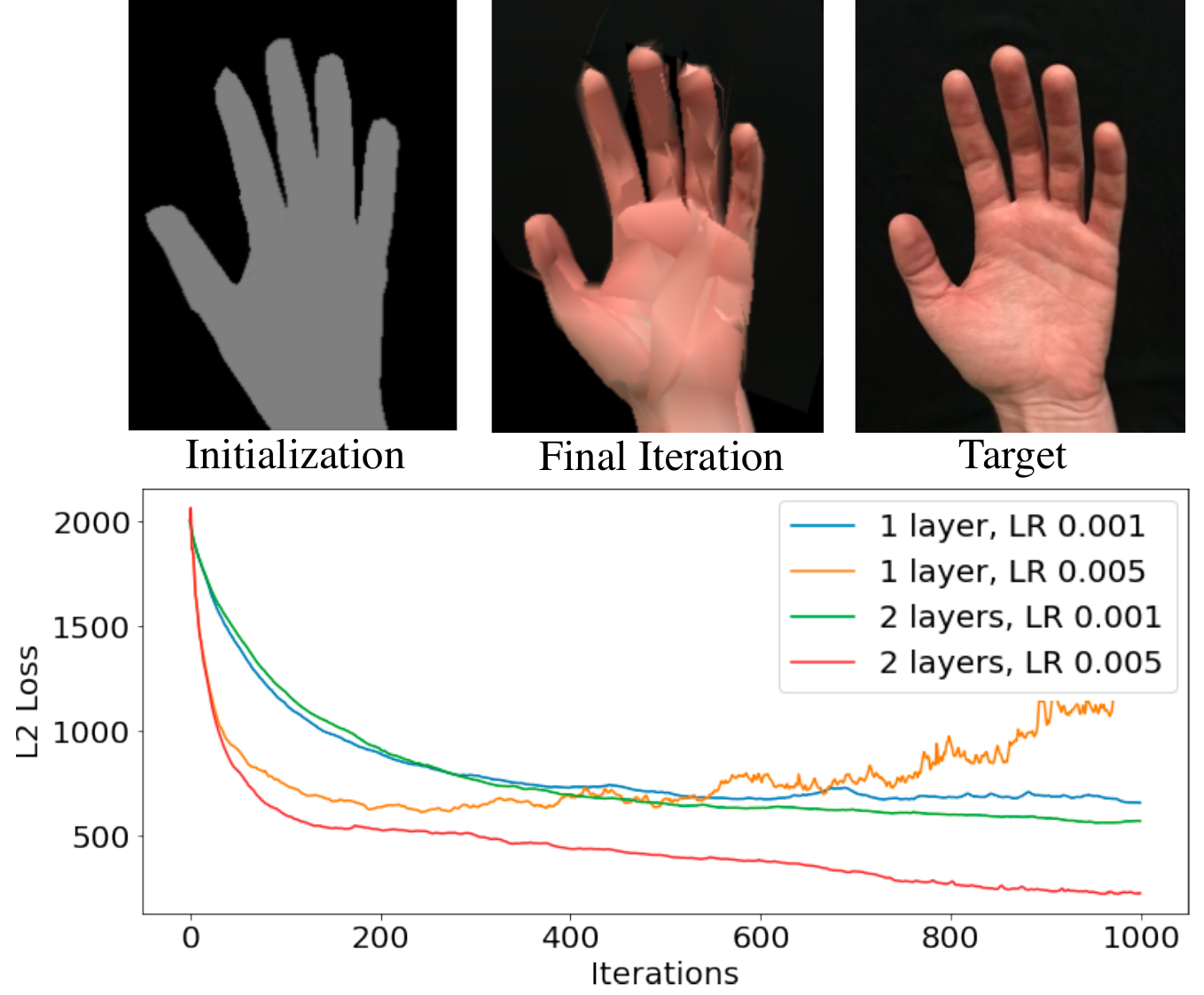}
    \caption{\textbf{Hand mesh optimization.} Given a roughly aligned template mesh (a), the vertex positions and colors are refined (b) to fit 3 target photographs (c, one shown). Because of the flexibility of the optimization, single-layer splatting can be unstable, and multi-layer splatting is necessary for consistent results (bottom).
    }
    \label{fig:hand_optimization}
    \vspace{-1em}
\end{figure}

Fig.~\ref{fig:hand_optimization} shows optimization of the vertex positions and colors of a hand template mesh~\cite{deodr}. The hand is first roughly aligned to the input images and given a uniform gray color (Fig.~\ref{fig:hand_optimization}a), then optimized to match the input photographs from 3 camera views using Adam. The surface is regularized using an As-Rigid-As-Possible deformation energy~\cite{laplacian}. Rather than set a solid background color, a solid rectangle is placed behind the hand to show the effect of multiple layers of geometry.

As shown in Fig.~\ref{fig:hand_optimization} bottom, multi-layer splatting is important for convergence when optimizing the vertex positions. Convergence is slower for single-layer splatting, and single-layer optimization becomes unstable at higher learning rates. Each vertex depends on fewer pixels compared to pose estimation (Fig.~\ref{fig:airplane_optimization}) and the shape is more flexible, so errors in the image derivatives are more significant. 

\subsection{Parametric Surface Optimization}
\label{sec:spline_results}

\begin{figure}[b]
    \centering
    \includegraphics[width=\columnwidth]{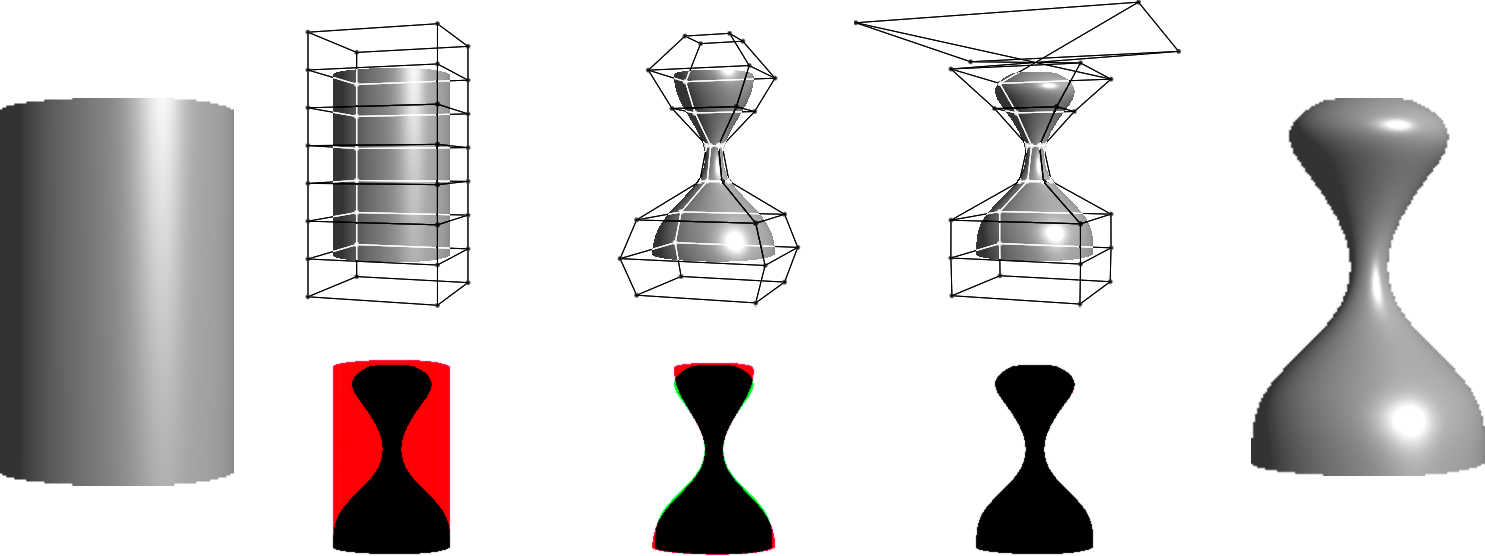}
    \caption{\textbf{Silhouette fitting of a B-spline surface.} A cylindrical subdivision surface (left) is deformed into a curved shape (right) by fitting silhouettes.  Top row shows the optimized control mesh and the corresponding subdivided surface.  Bottom row shows the silhouette overlap (black: silhouette agrees with ground truth, red: surface needs to be removed, green: surface needs to be added).}
    \label{fig:spline_results}
    \vspace{-0.5em}
\end{figure}

Fig.~\ref{fig:spline_results} shows a simple demonstration of silhouette optimization for a bicubic uniform B-spline surface~\cite{Catmull1978RecursivelyGB}. The surface is defined by a swept curve, producing an approximate surface-of-revolution. The variables of optimization are the radii of the 8 control points of the profile curve. The surface is initialized to a cylinder, and optimized to match the silhouette of another spline surface similar to a chess piece. The optimization converges in 200 iterations.

A triangle-based renderer would require differentiation through a densely tessellated mesh, whereas our method only uses tessellation to rasterize the surface parameters inside the non-differentiable sampling function $\mathcal{U}$. Once rasterized, the surface parameters are used to differentiably interpolate the control points, shade, and splat the silhouette image without tessellation. 

\subsection{Implicit Surface Optimization}

\begin{figure}[t]
    \centering
    \includegraphics[width=\columnwidth]{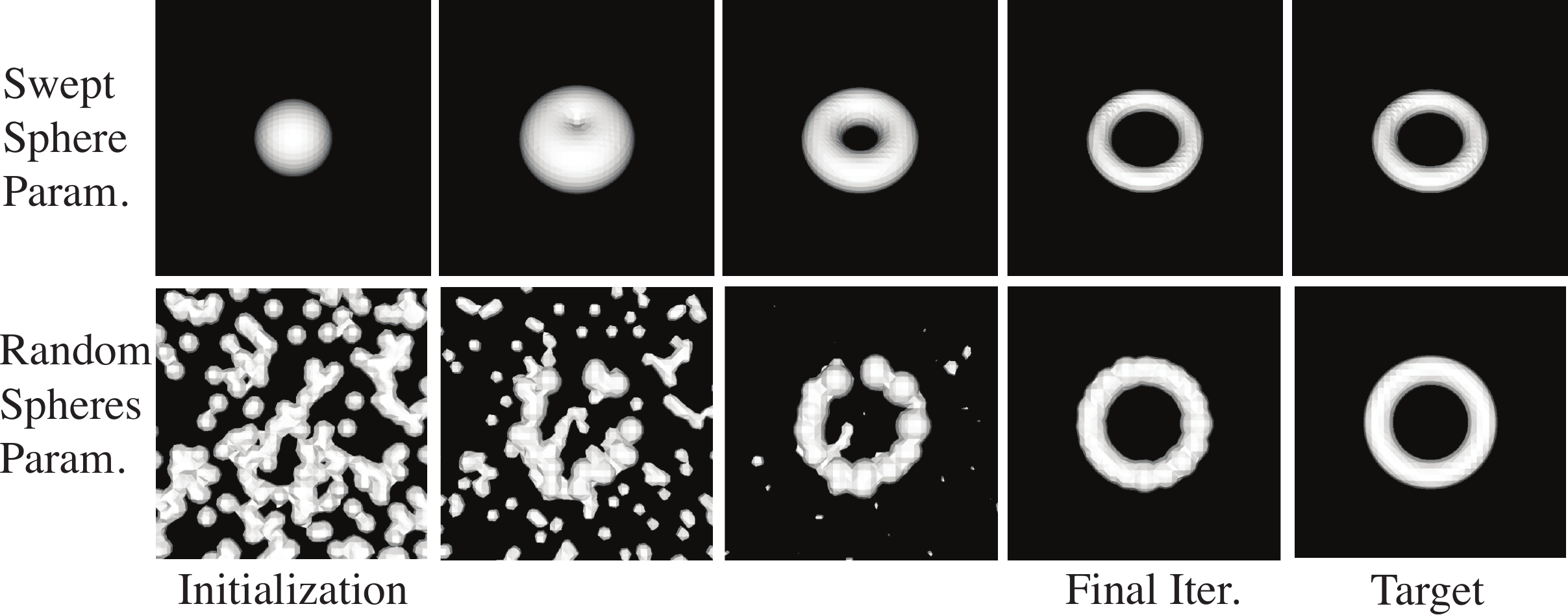}
    \caption{\textbf{Implicit surface optimization.} An SDF is defined on a $50^3$ grid using two different parameterizations: Swept Sphere (top) has ring radius and sphere radius parameters, while Random Spheres (bottom) has $n$ spheres each with 2-D center and radii parameters. Both are optimized to match an image of a torus.
    }
    \label{fig:sphere_to_torus}
\end{figure}

Fig.~\ref{fig:sphere_to_torus} shows fitting an image of a torus using an implicit surface and demonstrates that our method can handle topology changes. We show two possible parameterizations based on spheres: the first sweeps a sphere of radius $r_1$ along a circle of radius $r_2$, and a second defines the surface as the union of 200 spheres with individual radii and 2-D positions. The loss is mean-absolute error between the rendering and the target. Both optimizations are run for 400 iterations using Adam~\cite{adam}. Note that for the swept sphere initialized in the center of the torus, a rotated camera view (Fig.~\ref{fig:sphere_to_torus} top) is necessary to break symmetry that otherwise traps the optimization in a local minimum. 

These results may be compared to MeshSDF (Fig. 3 in \cite{meshsdf}), which also optimizes a sphere into a torus to demonstrate change in topology. In their case, however, the parameterization is a latent code of a deep neural network trained to produce spheres and tori. Unlike MeshSDF, our method does not rely on a deep neural network to compute gradients, so we are free to choose any implicit surface parameterization that can be evaluated on a grid.

\subsection{Surface NeRF}

\begin{figure*}[t]
    \centering
    \begin{tabular}{cccccc}
         \includegraphics[width=0.14\linewidth]{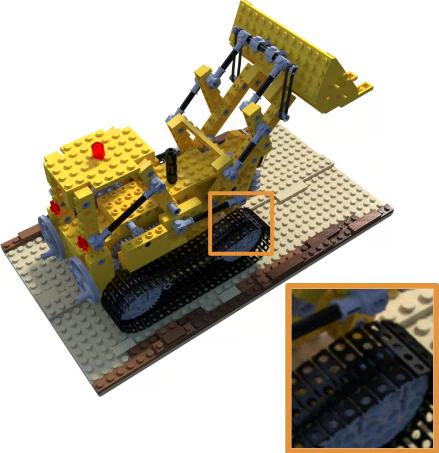} & 
         \includegraphics[width=0.14\linewidth]{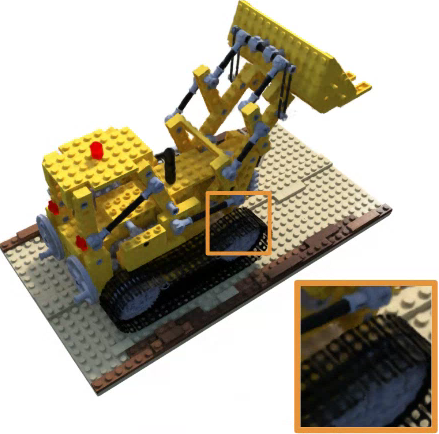} & 
         \includegraphics[width=0.14\linewidth]{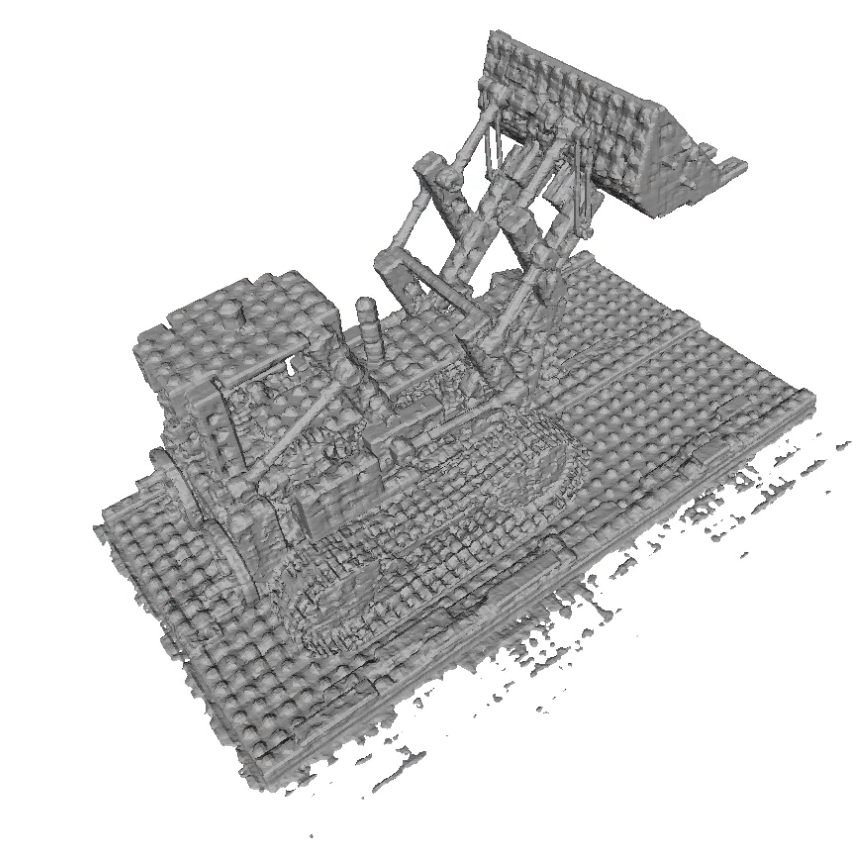} & 
         \includegraphics[width=0.14\linewidth]{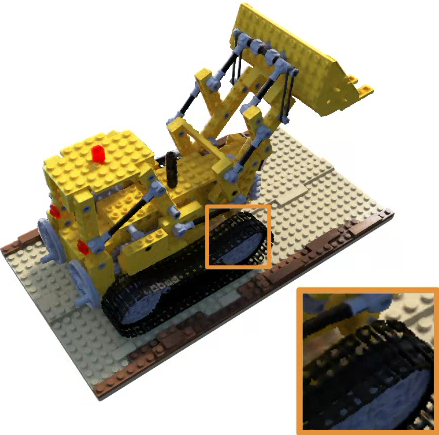} & 
         \includegraphics[width=0.14\linewidth]{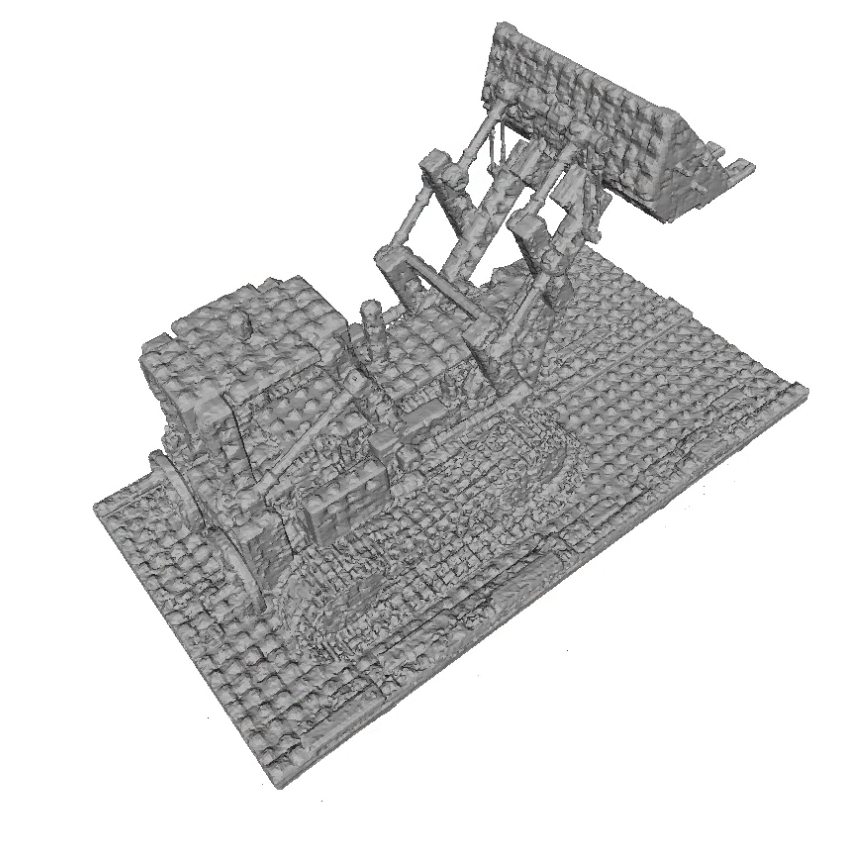} &
         \includegraphics[width=0.14\linewidth]{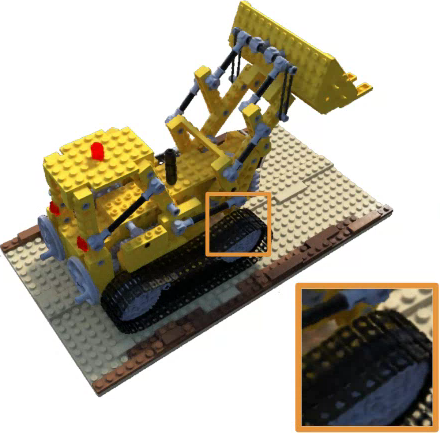}  \\
         \includegraphics[width=0.14\linewidth]{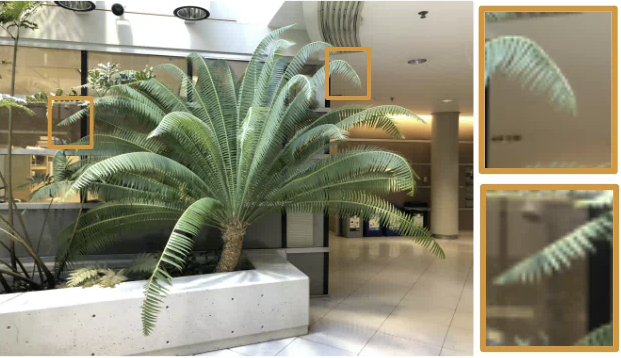} & 
         \includegraphics[width=0.14\linewidth]{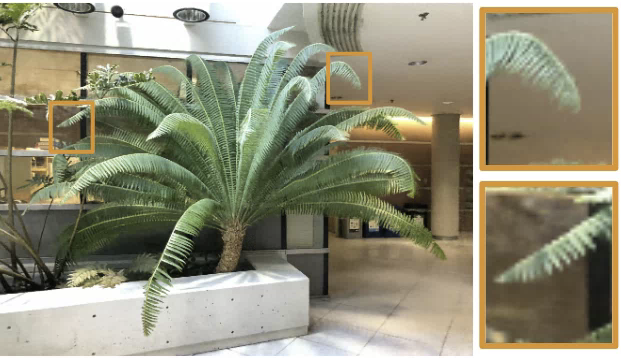} & 
         \includegraphics[width=0.14\linewidth]{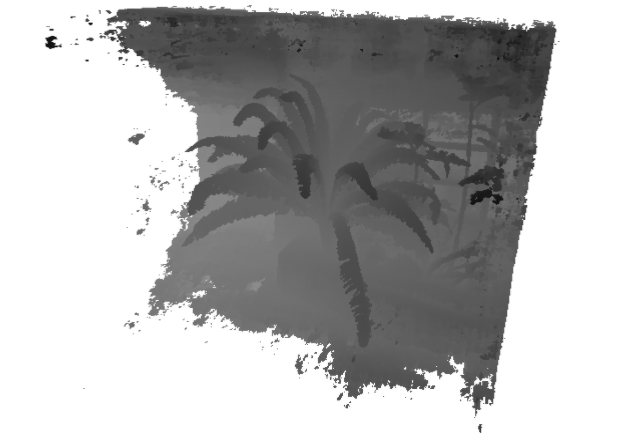} & 
         \includegraphics[width=0.14\linewidth]{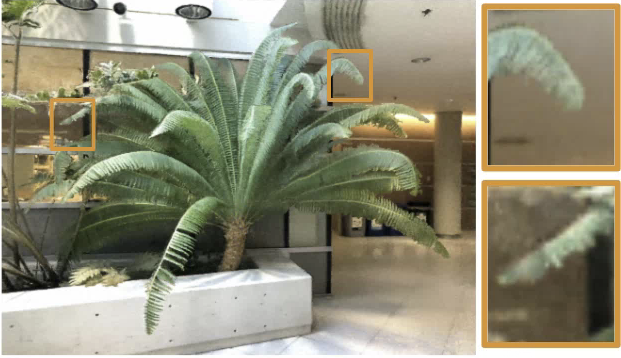} & 
         \includegraphics[width=0.14\linewidth]{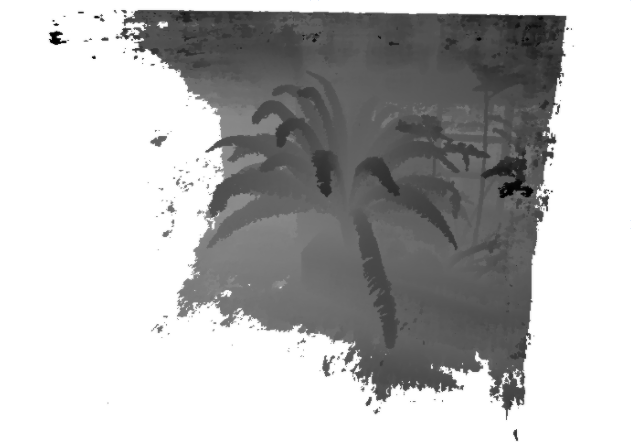} & 
         \includegraphics[width=0.14\linewidth]{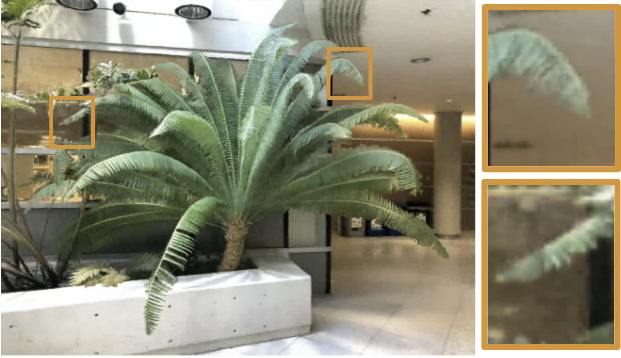} \\
         (a) Test Image & (b) NeRF\cite{nerf} & (c) NeRF surf. & (d) SNeRF (fixed) & (e) Opt. surf. & (f) SNeRF %
    \end{tabular}
    \caption{\textbf{Surface rendering results on NeRF test scenes.} Image quality is evaluated on held-out views (a). Full volume rendering of the NeRF produces a high-quality image (b), but an isosurface extracted from the trained NeRF (c) is noisy. With this surface fixed, the finetuned NeRF produces reduced quality (d). After optimizing the surface and finetuning the NeRF, we obtain an improved surface (e) and a high-quality rendering (f). %
    }
    \label{fig:nerf_results}
    \vspace{-0.5em}
\end{figure*}

NeRF~\cite{nerf} solves for a function from 3-D position $\mathbf{x}$ and viewing direction $\mathbf{d}$ to RGB color and density $\sigma$, such that when this function is rendered using volumetric raycasting, the result matches a set of posed input images. This method produces very high-quality view synthesis results, at the price of long rendering times. Using our approach, however, we can convert a pre-trained NeRF into a 3D mesh and a \emph{surface light field}~\cite{slf}. This representation requires only a single NeRF evaluation per pixel vs. the 128 required by volume rendering, reducing per-pixel cost from 226 MFLOPS to 1.7 MFLOPS. 

\noindent\textbf{Surface Optimization.}
We first evaluate NeRF on a regular grid to construct the input density field, then extract an isosurface at a fixed threshold. This surface is an accurate but noisy model of the subject shape (Fig.~\ref{fig:nerf_results}c). Furthermore, since NeRF is trained under the assumption of volume rendering, it must be finetuned in order to produce good results when evaluated only at the surface.

Given this initialization, we directly optimize the density grid while finetuning the NeRF network under the original NeRF L2 loss. We render the isosurface using RtS, where the NeRF network produces the shaded colors $\mathbf{C}_k$, effectively acting as a surface light field~\cite{miller98}. After optimization, the isosurface is refined (Fig.~\ref{fig:nerf_results}e), and the output RGB has similar quality to the original NeRF rendering (Fig.~\ref{fig:nerf_results}f). We use Adam~\cite{adam} for finetuning the NeRF network and standard gradient descent for optimizing the density grid, as Adam is unstable for grid optimization due to the isosurface not constraining all grid values at each iteration. At each iteration, we take one gradient step on the NeRF network while holding the density grid fixed, followed by one gradient step on the density grid while holding the NeRF network fixed. We use an isosurface threshold of 50 (suggested by \cite{nerf}) and optimize for 5000 iterations.

\begin{table}[b]
    \centering
    \begin{tabular}{l|cccc}
                            & \multicolumn{2}{c}{Lego} & \multicolumn{2}{c}{Fern}\\
                            & PSNR & SSIM & PSNR & SSIM \\\hline
         NeRF (baseline)    & 29.91 & 0.962 & 24.38 & 0.864 \\
         SNeRF              & 29.61 & 0.957 & 23.52 & 0.813 \\
         SNeRF (fixed)      & 27.44  & 0.945 & 23.41 & 0.809 \\
         
    \end{tabular}
    \caption{\textbf{Image quality for NeRF methods.} PSNR and SSIM for two scenes for baseline NeRF~\cite{nerf}, our surface NeRF (SNeRF), and SNeRF without geometry optimization (fixed surface).}
    \label{tab:nerf_table}
\end{table}

\noindent\textbf{Evaluation.} Table~\ref{tab:nerf_table} shows the results of our approach on the NeRF ``Lego'' and ``Fern'' datasets (the two for which pre-trained weights are available). We compare image quality for baseline NeRF and our Surface NeRF (SNeRF). We also evaluate SNeRF without surface optimization (``fixed surface'') where the NeRF network is finetuned by only sampling on the fixed isosurface. On Lego, a scene that was synthesized from a surface, SNeRF achieves PSNR within $0.3$ of NeRF and improves $2.2$ PSNR over the baseline without surface optimization. On Fern, SNeRF loses $0.9$ PSNR to full volume rendering, and improves only $0.1$ PSNR over the fixed surface baseline. This result is likely due to the extremely complex isosurface of the Fern scene (Fig.~\ref{fig:nerf_results}).

\section{Discussion and Limitations}

Rasterize-then-splat is a general method that addresses two key issues of differentiable surface rendering: handling varying surface representations, and providing derivatives at occlusion boundaries. Our method applies to any surface that can be expressed as a non-differentiable \emph{sampling} and a differentiable \emph{evaluation} function. This flexibility opens the door for researchers to explore surface representations not previously supported by differentiable rendering, including spline surfaces and general isosurfaces. We have demonstrated that isosurface rendering can be used to reduce the runtime cost of NeRF rendering by more than $100\times$. 

Our method requires a closed-form evaluation function, which may not be available at all (e.g., some subdivision schemes) or only available via a Marching Cubes discretization. For surfaces that are defined as continuous functions of space, the discretization can affect surface quality. Since the evaluation happens only near the surface, however, quality may be improved by increasing resolution at a quadratic (not cubic) cost in evaluations.

While we render multiple layers in order to resolve occlusions, the splatting step currently assumes a single surface at each pixel and does not handle semi-transparent objects. A direction for future work is to extend the method to handle semi-transparent layers, which could improve quality on scenes that include reflections or translucency.

{\small
\bibliographystyle{ieee_fullname}
\bibliography{references}
}

\end{document}